\documentclass[preprint2]{aastex}

\newcommand{\etal}{et~al.~}

\shortauthors{W. -K. Park \etal}
\shorttitle{CQUEAN}

\slugcomment{Accepted for publication in PASP}

\begin{document}

\title{Camera for QUasars in EArly uNiverse (CQUEAN)}

\author{Won-Kee Park$^{1,2}$}
\affil{$^1$ CEOU/Department of Physics and Astronomy, Seoul National University,
Seoul 151-742, Korea}
\affil{$^2$ Korea Astronomy and Space Science Institute, Daejeon 305-348, Korea}
\email{wkpark@astro.snu.ac.kr}
\author{Soojong Pak$^{3,4}$}
\affil{$^3$ School of Space Research, Kyung Hee University, Gyeonggi-Do 446-741, Korea}
\affil{$^4$ Department of Astronomy, University of Texas at Austin, Austin, TX 78717}
\email{soojong@khu.ac.kr}
\author{Myungshin Im$^1$, Changsu Choi$^1$, Yiseul Jeon$^1$}
\affil{$^1$ CEOU/Department of Physics and Astronomy, Seoul National University,
Seoul 151-742, Korea}
\email{mim@astro.snu.ac.kr, changsu@astro.snu.ac.kr, ysjeon@astro.snu.ac.kr}
\author{Seunghyuk Chang}
\email{chang@offaxis.co.kr}
\author{Hyeonju Jeong$^3$, Juhee Lim$^3$, \and Eunbin Kim$^3$}
\affil{$^3$ School of Space Research, Kyung Hee University, Gyeonggi-Do 446-741, Korea}
\email{jhyeonju@khu.ac.kr, juheelim@khu.ac.kr, ebkim@khu.ac.kr}

\begin{abstract}
 We describe the overall characteristics and the performance
 of an optical CCD camera system, Camera for QUasars in EArly uNiverse (CQUEAN), which is being used
 at the 2.1 m Otto Struve Telescope of the McDonald Observatory since 2010 August.
 CQUEAN was developed for follow-up imaging observations of red sources
 such as high redshift quasar candidates ($\mathrm{z} \gtrsim 5$), Gamma Ray Bursts, brown dwarfs,
 and young stellar objects. For efficient observations of the red objects, CQUEAN
 has a science camera with a deep depletion CCD chip which boasts a higher quantum efficiency
 at $0.7 - 1.1 \, \mu m$ than conventional CCD chips.
 The camera was developed in a short time scale ($\sim$one year), and has been working reliably.
 By employing an auto-guiding system and a focal reducer to enhance the field of view
 on the classical Cassegrain focus, we achieve a stable guiding in 20 minute exposures,
 an imaging quality with FWHM $\geq 0.6\arcsec$ over the whole field ($4.8\arcmin \times 4.8\arcmin$),
 and a limiting magnitude of $z = 23.4$ AB mag at 5-$\sigma$ with one hour total integration
 time.
\end{abstract}



\section{INTRODUCTION}


Recent emphasis on the distant and/or obscured target 
observation requires near infrared (NIR) sensitive imaging devices. Although the conventional
 CCD chips, especially the back illuminated, thinned CCD chips,
 are very sensitive over a wide wavelength range
 ($0.4 - 0.8 \, \mu m$), most of them are not sensitive
 in NIR bands, with the quantum efficiency (QE) of order of
 a few percent at $\sim 1 \, \mu m$ \citep[e.g.,][]{im10a}.
 Detectors made of low band-gap semiconductor
materials such as HgCdTe and InSb offer high sensitivity in NIR,
but the NIR arrays are expensive in cost, and have
strict export restrictions, making them less accessible to many
astronomical imaging applications. Considering the recent
high demand for NIR imaging, it is thus important to develop
a cost-effective imaging camera.

 To facilitate imaging out to $1.1 \, \mu m$ at a reasonable cost
and within a fast development schedule,
 the Center for the Exploration of the Origin
 of the Universe (CEOU) has developed a CCD camera system,
 Camera for QUasars in EArly uNiverse (CQUEAN). It is
a camera system designed primarily for the observation of high redshift
quasar candidates.
 The design goal was to make a camera sensitive at
 $0.7 - 1.1 \, \mu m$ (QE $>$ 20 \% at $1 \, \mu m$)
 with a moderate field of view (at least a few arcmin
 across). The sensitivity in the red is governed by the need
 for the detection of the Lyman break of objects at $\mathrm{z} \gtrsim 5$.
 The requirement for the field of view is to allow
 photometry based on the reference stars available from
 large area surveys such as the Sloan Digital
 Sky Survey (SDSS) \citep{yor00} and the Large Area Survey of UKIRT Infrared
 Deep Sky Survey (UKIDSS)  \citep{law07}.
 The optical performance requirement was to obtain 
 the seeing limited point spread function (PSF) with FWHM better than $1\arcsec$.

Based on these requirements, we adopted a CCD camera with a deep depletion chip, a focal reducer, and an auto-guiding system. The development of CQUEAN started in the summer of 2009, and it was installed on the Cassegrain focus of the 2.1 m Otto Struve  telescope at the McDonald Observatory, Texas, USA, on 2010 August 12. 
Since its commissioning run, CQUEAN has been used for various scientific projects.

 In this paper, we describe the overall characteristics of CQUEAN, and
 its performance based on the first year of the scientific operation.
 Section 2 describes the overview of the instruments including the science 
 camera, its filters, and the parameters of CQUEAN on the 2.1 m telescope. In Section 3, we describe
 various characteristics of the camera and images obtained with CQUEAN.
 We report the performance of CQUEAN
  in Section 4, and briefly describe the scientific result obtained with CQUEAN
 in Section 5. Finally, summary and conclusion are
 presented in Section 6.

\section{INSTRUMENT DESCRIPTION}

\begin{figure*}
\epsscale{1.5}
\plotone{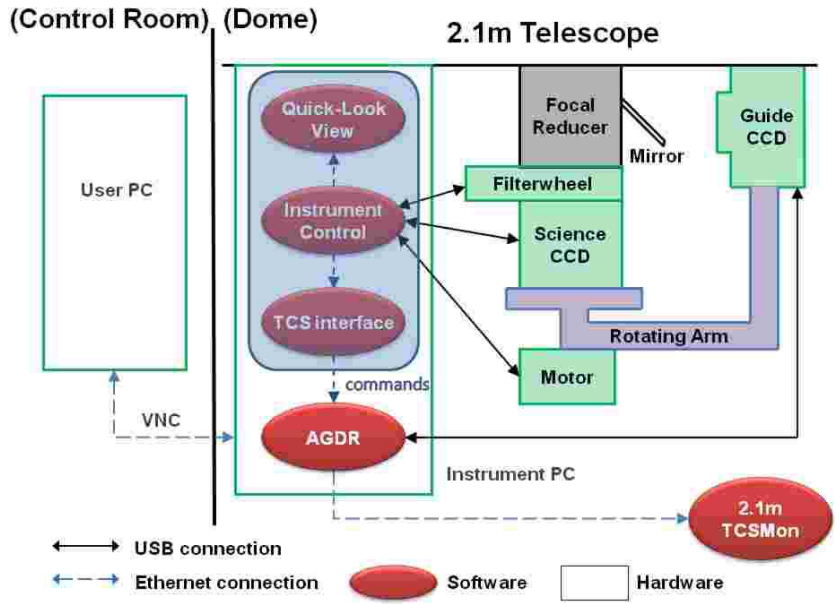}
\caption{System architecture of the CQUEAN system \citep{kim11}.
The ellipses and the squares denote the software and the
hardware parts, respectively. The arrows indicate the directions of
command and data flows. All units of CQUEAN including the control computer
are attached to the telescope Cassegrain focus. The observers access the 
control computer from the control room via an ethernet network.}
\label{structure}
\end{figure*}

  The main units of CQUEAN are: a science camera,
  a filter wheel that houses seven filters, a focal reducer, and an auto-guide camera.
  The auto-guide camera is mounted on a rotating arm to secure a large off-axis
 field of view \citep{kim11}. Fig. \ref{structure} shows the overall structure of CQUEAN system.
 An enclosure 
holding the science
 camera system, the guide camera system as well as a motor to move the guide
camera were made with several pieces of aluminum. 
 The whole system is operated by a custom-made software package on a 
 linux-running computer which is mounted next to the main units.
 The observers access the control computer over network during observations.
We describe each unit in detail below.

\subsection{Science camera}
CQUEAN uses an iKon-M M934 BR-DD model from Andor Technology PLC\footnote{http://www.andor.com},
as its main science camera. 
 It uses an E2V $1024 \times 1024$ pixel deep-depletion CCD as its detector.
 This CCD shows a higher red sensitivity than
conventional back-illuminated thinned CCDs, and also shows little fringe pattern as described
later.
In addition to the CCD chip, the camera contains a readout electronics, and the cooling
unit inside the camera body. A fused silica window and a mechanical shutter are
installed in front of the CCD chip.
The camera is connected to the control computer via USB port, which makes it easy for
us to handle but limits the cable length between the camera body and the computer.

A thermoelectric cooler is used in the camera, which can cool down the chip to  $-80\,^{\circ}$C.
However, after reviewing performance test from the chip manufacturer,
 we decided to operate the camera at $-70\,^{\circ}$C 
 because the higher operating temperature improves
QE at longer wavelengths. Even at this higher operating temperature,
the dark current is still comparable to the sky background.

The pixel readout speed and preamp gain can be set to
three different values, respectively. 
After several tests, we decided to operate the
camera with the pixel readout speed of 1 MHz, and with the preamp gain of 1. 
Under these settings, it takes about one second to read the whole pixels. 
 A summary of the technical specification of the camera is listed
 in Table \ref{ccdspec}. 

\begin{figure*}
\epsscale{1.0}
\plotone{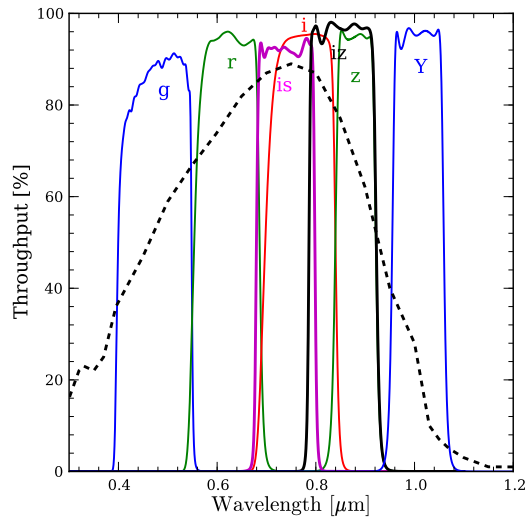}
\caption{QE of the science camera chip and the throughputs of the CQUEAN filters.
CCD QE is shown with black dashed line, while throughputs of filters are shown 
with solid lines.
}
\label{throughput}
\end{figure*}

A generic curve of the QE of the chip at $-100\,^{\circ}$C is shown in Fig. \ref{throughput}
 with a thick dashed line, which is taken from the manufacturer's brochure.
The characteristics of deep depletion E2V chips suggests that the QE increase by 25 \% 
at $1\,\mu m$ when the chip temperature is warmer by $30\,^{\circ}$C at the expense of 
a reduced sensitivity at the blue end according to the manufacturer.
 Since the operating temperature for CQUEAN was set to $-70\,^{\circ}$C, the
 actual QE at long wavelength is 
 higher by a factor of 1.2 than shown in the figure.

\subsection{Filters}
In front of the science camera, we installed a commercial filter wheel, 
CFW-1-8, made by the Finger Lake Instrumentation (FLI)\footnote{http://www.flicamera.com},
USA. This wheel houses eight circular filters with a diameter of 28 mm and a thickness of 5 mm.

We installed seven filters, and one black anodized aluminum plate for a calibration 
purpose.
 The seven filters are $g$, $r$, $i$, $z$, $Y$, $is$, and $iz$, which were manufactured by
 Asahi Spectra Co., Ltd.\footnote{http://www.asahi-spectra.co.jp}
 The $g, r$, and $i$ filters are very similar to those for SDSS \citep{fuk96},
 while our $z$ filter has a transmission curve identical to that of the LSST Z-filter,
which cuts off photons beyond $0.92 \, \mu m$.
The $Y$-band filter is identical to LSST Y-band filter whose bandwidth is
narrower than that described in \citet{hil02} in order to minimize the 
contamination by telluric OH emission lines.
The custom designed $is$ and $iz$ filters have bandwidths of $\sim 100$ \AA,
and their passbands are located between $r$ and $i$, and $i$ and $z$, respectively. They are designed to
detect the quasar candidates at redshift of $5.0 < \mathrm{z} < 5.7$, since SDSS $r$, $i$ and $z$ filter
 colors  have difficulty in distinguishing these quasars from the Galactic cool stars. 

Fig. \ref{throughput} shows the transmission curves of our filters as well as the QE of
the CCD. 
It can be seen that CQUEAN does have an excellent throughput
even at $Y$-band range, compared to other typical optical CCDs.

\subsection{Focal reducer\label{freducer}}

 The 2.1 m telescope has a focal ratio of $f/13.65$ at its classical Cassegrain
focus. 
This large focal ratio would result in 
the field of view of $1.6\arcmin \times 1.6 \arcmin$ for our
science camera without a re-imaging
optics, which is not adequate for our scientific
 purposes. 
 The pixel scale of $0.094 \arcsec$/pix is also too small for 
 the seeing condition at the McDonald Observatory which is $0.7\arcsec - 1.2 \arcsec$ in $V$-band. 

We made a custom focal reducer to enlarge the field of view as well as to 
correct optical aberrations,
 especially for coma. Considering 
the seeing size, 
it is designed to reduce the effective focal length by three fold. In this 
configuration, the corresponding pixel scale
is  $0.28 \arcsec$/pix, so that 3 -- 4 pixels can cover the FWHM of a 
stellar profile under a typical seeing condition. The field of view of the system becomes
$4.8\arcmin \times 4.8\arcmin$.

\begin{figure*}
\epsscale{1.2}
\plotone{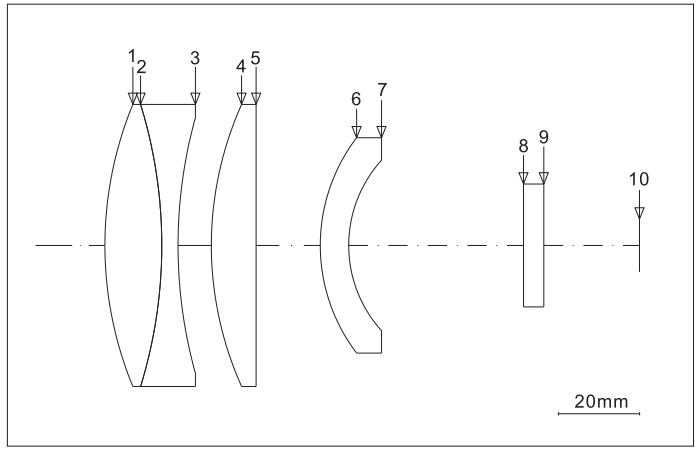}
\caption{Optical layout of the focal reducer. See Table~\ref{tbl2} for the descriptions
of the optical elements.}
\label{lens}
\end{figure*}

\begin{figure*}
\epsscale{1.3}
\plotone{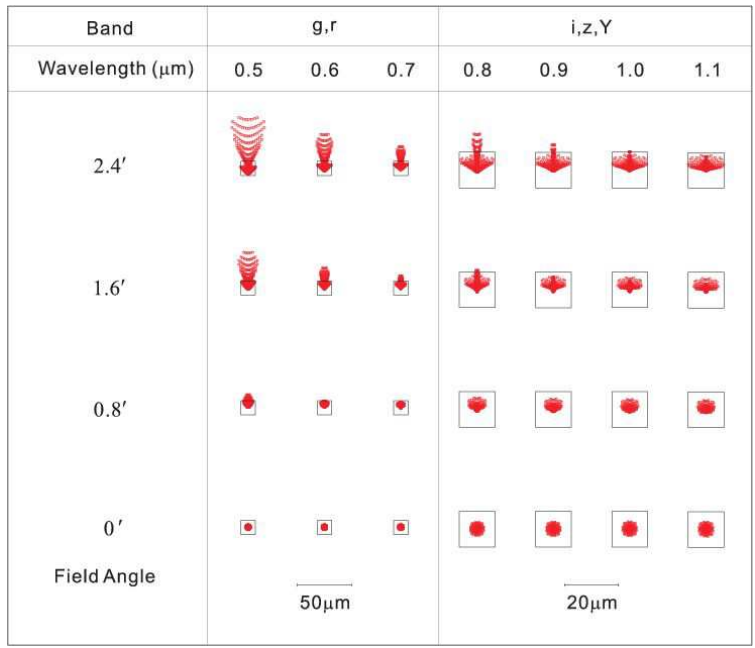}
\caption{Spot diagrams for each wavelength band at its focal plane position. The sizes of
the square boxes are the same with that of a single CCD pixel.}
\label{spot}
\end{figure*}

In order to reduce the optical complexity, we did not seriously consider the
chromatic aberration.
The focal reducer consists of four lenses and is optimized for the wavelengths
between 0.8 and $1.1 \,\mu m$. Fig.~\ref{lens} shows the side view of the reducer optics.
The first three elements reduce the focal length while correcting coma aberration of the
telescope. The last element corrects the residual astigmatism and field curvature.
A summary of the reducer optics is given in Table~\ref{tbl2}. Also,
the spot diagrams for each wavelength band are shown in Fig.~\ref{spot}. All air to
glass surfaces are coated to reduce the reflections for incident light of $0.8-1.1 \,\mu m$
wavelength band \citep{lim12}. Reflections for the light outside this wavelength band
were not considered since the optical performance of the reducer was optimized for
$0.8-1.1 \,\mu m$ band only. Therefore, the anti-reflection (AR) coating substantially reduces
the throughput for wavelength bands shorter than $i$-band. The optical
components in the focal reducer was manufactured by Green Optics Co., Ltd.\footnote{http://www.greenopt.com}

 Analysis of the test images obtained during the engineering run reveals that the 
focal reducer indeed works as 
expected: the images in $i$, $z$, and $Y$-bands are perfect within the seeing limits,
while those in $g$ and $r$-bands show coma aberrations on the peripheries of the fields.
Because we did not consider the fields at four diagonal corners of the field of view,
some vignetting pattern arises at all corners. 
This vignetting can be reduced by flat-fielding to a certain degree, but
four corners of the chip are not recommended for scientific use. Section \ref{section_psf} describes
the characteristics of the image profile of the CQUEAN optical system. Detailed description
on the fabrication of the focal reducer and its performance is given by \citet{lim12}.

\subsection{Guide camera}

\begin{figure*}
\epsscale{1.4}
\plotone{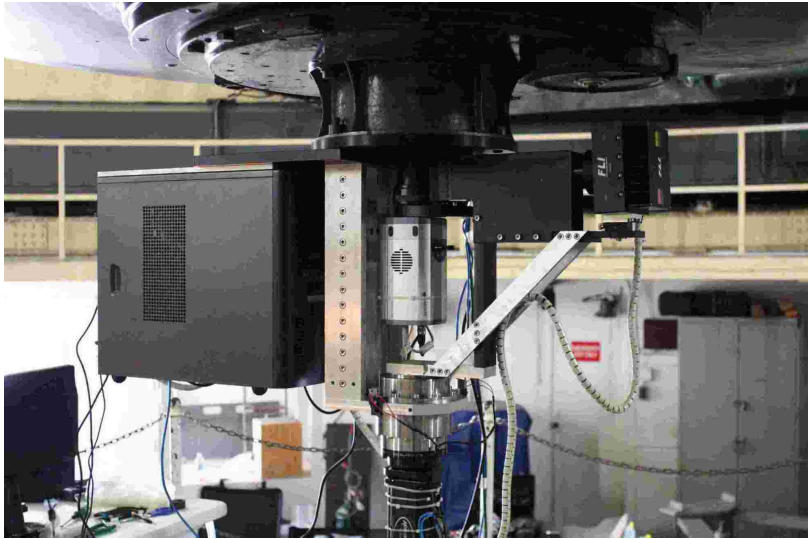}
\caption{CQUEAN system attached to the Cassegrain focus of 2.1 m telescope. The black box on the
left is the control computer, and a small silver box at the center is the science camera. Motor
below the science camera controls the guide camera arm that holds the guide camera, a small
black box on the right part of the photo.}
\label{sysphoto}
\end{figure*}

 As described in \citet{kim11}, the 2.1 m telescope shows 
tracking errors which can be removed by feedback correction in every few seconds.
We made 
an auto-guiding system for CQUEAN and the telescope 
by installing an off-axis guide camera. 
ProLine 1001E model from the Finger Lakes Instrumentation
was used as a guide CCD camera. It has a $1024\times1024$ pixel CCD chip with 
pixel size of $24\,\mu m \times 24\,\mu m$. It is cooled
to $-30\,^{\circ}$C with thermoelectric cooling unit during the operation. An off-axis pick-up
mirror is installed to feed an off-axis
field of the 2.1 m telescope to the guide camera. There is no additional optics for the guide
camera, except a baffle to block the stray light into the optical path between the mirror
and the guide camera. Direct imaging of an off-axis field results in the pixel scale of
$0.174\arcsec$/pixel, and the total field of view of $2.97\arcmin \times 2.97\arcmin$. The expected
number of guide stars brighter than limiting magnitude for one second exposure
on the field of view is about one in the fields of galactic pole region,
 which may make auto-guiding impossible due to a lack of adequate guide stars. Therefore
we use a rotating arm for the whole guide camera system to increase the size of
the viewable off-axis field by a factor of five. 
The arm assembly is moved with a stepping motor located
behind the science camera (see Fig.~\ref{sysphoto}).
Both the guide camera and the moving mechanism are controlled with the same computer.

\subsection{Operating software}

 The observers need to control several components of CQUEAN to obtain the
imaging data. They also need a quick and efficient way to check the quality
of the acquired data on the spot. Therefore, we developed a GUI program
for CQUEAN users which is easy to use, while providing functionalities
sufficient to acquire the scientific data.

The program is written in Python version 2.6.5 for easy development and maintenance.
The front-end GUI is developed in-house with \verb|wxPython| library, while the low-level
instrument control libraries are supplied by vendors of the science camera and
the filter wheel. To access C-based SDK libraries with python, \verb|ctypes| package was used.
In addition to those, \verb|NumPy| (math library) and \verb|matplotlib| (graphic library)
are required for the mathematical analysis routines used in the software, and \verb|PySerial|
for serial communication with the motor for the guide camera arm.

\begin{figure*}
\epsscale{1.4}
\plotone{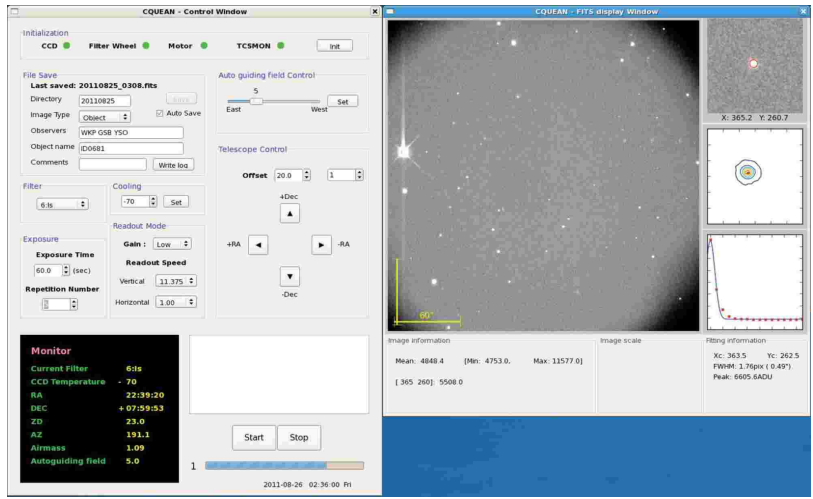}
\caption{Screenshot of CQUEAN control program running on a linux machine. The left window
is the CQUEAN Control window, and the right one Quick-Look window.}
\label{gui}
\end{figure*}

The control program runs on a linux computer running CentOS 5.5, and it requires
X Window System for displaying the GUI components of the program. Its
GUI consists of two windows: the CQUEAN Control window and the Quick-Look window.
With the CQUEAN Control window, one can control all the CQUEAN components to acquire the
imaging data: the science camera, the filter wheel, and the guide camera moving mechanism.
It displays the current status of these components. In addition, it interacts with
Telescope Control System (TCS) of the telescope to obtain the current pointing information of the telescope.
The Quick-Look window displays the acquired image data on the screen. It has quick
analysis features of displaying the contour/radial profile of a source
selected with a click on the displayed image. Fig. \ref{gui} shows the two
windows of CQUEAN control program.
The imaging data are saved with 32 bit integer values to a FITS format file
\citep{wel81}
after adding various header keywords to facilitate data reduction.

The auto-guiding software, \textit{agdr}, runs on the same linux PC that runs
the CQUEAN control program. It has been used for the auto-guiding of the 2.7 m telescope
at the McDonald Observatory. A modified version for CQUEAN was supplied by P. Sam
Odoms of the McDonald
Observatory. The \textit{agdr} controls the guide camera to obtain images of
a guide star with  1 -- 5 second exposures, depending on the brightness of the
guide stars. 
It determines the position of the guide star on a frame, and calculates offsets
 of the position with respect
to the position of the guide star on a reference frame. Then the \textit{agdr} interacts with 
TCS of the 2.1 m telescope to correct the drift based on the offset obtained.
The program repeats the procedures until the data acquisition with the science camera is completed.
The auto-guiding performance was tested up to 1200 seconds of exposure. The test shows 
that auto-guiding with 1200 second exposure does not introduce any serious tracking 
errors \citep{kim11}. We expect that longer exposures are possible without much trouble, 
although we expect that most users use exposures not much longer than 1200 seconds.
 
 The \textit{agdr} software is also useful for monitoring the seeing condition
and the atmospheric transparency of the night, since it keeps updating the FWHM
and instrumental magnitude of the guide star on a plot.

\section{CHARACTERISTICS OF CQUEAN}

CQUEAN saw its first light during the engineering run carried out at the McDonald
Observatory from 2010 August 10 to 17. Fig.~\ref{sysphoto} shows CQUEAN
attached to the 2.1 m telescope.
During the engineering run, we calibrated 
our guide
camera system so that it can auto-guide the telescope correctly.
Bias and dark images of the science camera were taken under several different 
temperatures to check the temperature dependencies,
though most of subsequent calibration and scientific
observations were carried out with the temperature set to $-70\,^{\circ}$C. Various readout modes 
were tried to determine the best mode for scientific
observation in consideration of both the time overhead and the readout noise level. Several twilight
flats and dome flats were obtained, and we also investigated science camera linearity range
 and calibration procedure.

\subsection{Bias}

Bias images were taken to examine the stability in the DC offset during the observing nights.
They were obtained during the beginning
and the end of each night, and also just before and after observations of
each science target. 

\begin{figure*}
\epsscale{1.0}
\plotone{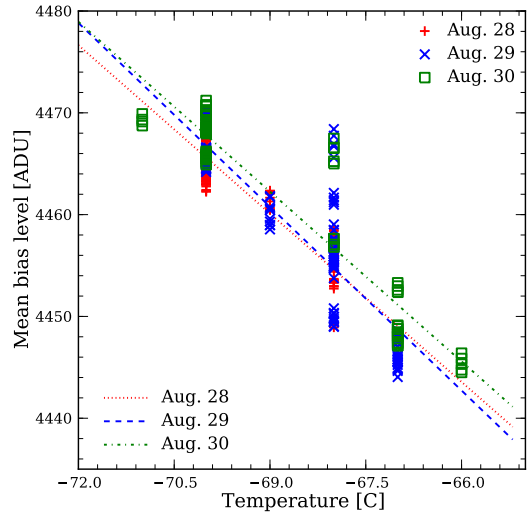}
\caption{Bias level of CQUEAN during three nights of 2011 August run against the chip
temperature. The each point in this plot represents the mean pixel value of a single bias
image. Bias images were obtained usually before and after the observation of a target.
The three lines indicate the fitted relation to samples from each night.}
\label{biaschange}
\end{figure*}

 The examination of bias images revealed that
 no particular patterns are seen in a single bias image, and pixel values are
uniform all over the region with a standard deviation of $\sim 8.2$ ADU
which is basically the readout noise.
 However, the bias level shows a small amount of inter- and intra-night
 variability, which may originate from the temperature variation of the chip or
 from the electrical instability. 
 Fig.~\ref{biaschange} shows the mean pixel value of single bias images
 during three nights in 2011 August run against the chip temperature.
 The bias level fluctuated during the run from about 4444 up to 4472 ADUs.
 Although there are scatters among the samples, and differences among
 each night, the bias level turned out to increase as the chip temperature
 gets lower.
%
The chip temperature is controlled by turning on and off the thermoelectric
cooler, and the frequency of the operation depends on the ambient temperature.
Therefore, we suspect that the temperature-sensitive variations in the bias
level is originating from the temperature-sensitive electronic components of
the camera.
%
Fig. \ref{biaschange} shows that the one degree change in temperature resulted in 
$\sim 6$ ADU in the mean bias level, although the change is not entirely predictable.
Sort-term variation over one hour timespan is less than $\sim 6$ ADU, so the interim
solution for the bias level fluctuation would be to use bias images taken right
before and after the observation of a target.
In future, we hope to introduce an overscan
area to correct for the bias level variation. For now, we suggest to take bias
images before and after the observation of a target. Since it takes only one second
 to read a bias image, this does not introduce much overhead.

\subsection{Dark}

\begin{figure*}
\epsscale{1.0}
\plotone{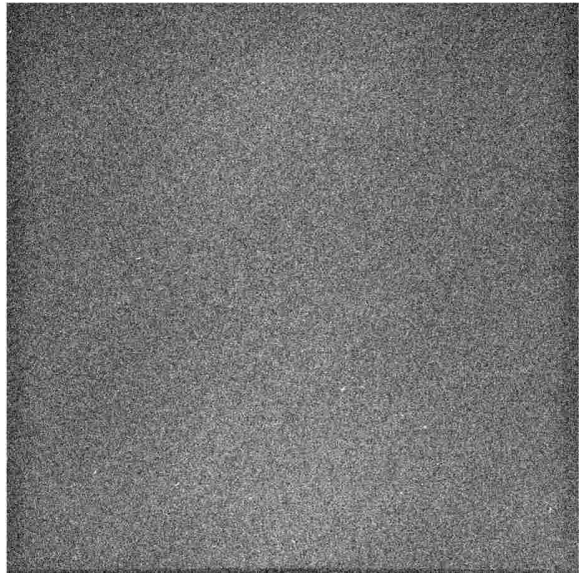}
\caption{Example of bias subtracted 300 second dark images of CQUEAN.}
\label{exdark}
\end{figure*}

 To correct for the dark current signal, the dark images are usually taken before the start
 or at the end of observation every night, with exposure times equivalent to those of science images.
 Fig. \ref{exdark} shows an example of bias subtracted dark images taken with CQUEAN with
 300 second exposure.
 The dark level is slightly higher
in central columns than at the edges. But except for this feature, dark image shows a relatively uniform
pixel value distribution over the image. The mean dark level was measured to be
$0.23 \pm 0.01$ electrons/sec/pixel, with the CCD chip at 
$-70\,^{\circ}$C.

\begin{figure*}
\epsscale{1.0}
\plotone{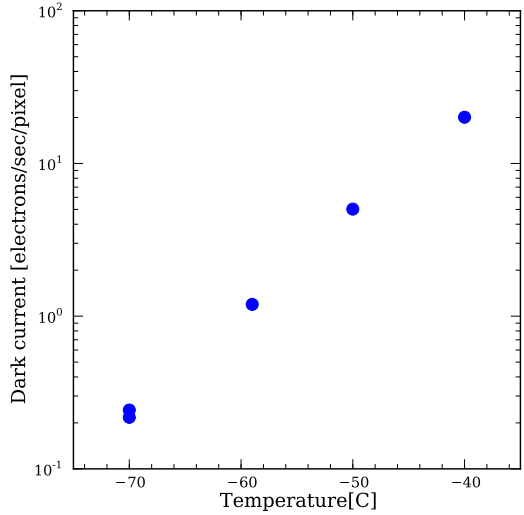}
\caption{Dark current levels of science camera against the cooling temperature.
 The plot indicates that the dark current increases by a factor of ten when
 the temperature increases by $10\,^{\circ}$C.}
\label{darktemp}
\end{figure*}

During a night in the 2010 August run, 
we obtained dark images under various temperatures: 
$-40\,^{\circ}$C, $-50\,^{\circ}$C, $-60\,^{\circ}$C,
and $-70\,^{\circ}$C. Fig. \ref{darktemp} shows the result of our experiment.
  The dark level increases exponentially as a function of the operating
 temperature.
 The operating temperature is optimized
 so that the camera provides a good throughput at $1 \, \mu m$ while keeping the
 dark current at a level significantly lower than the sky background.
  The sky count was found to be the lowest at $\sim 1$ ADU/sec/pixel in $g$-band, and
 3.8 ADU/sec/pixel in $Y$-band. 

\begin{figure*}
\epsscale{1.0}
\plotone{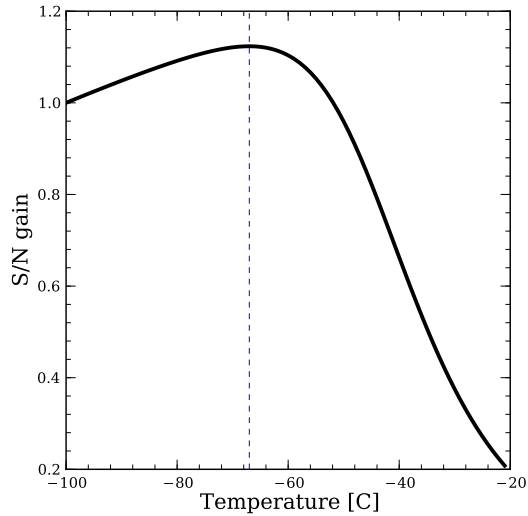}
\caption{
Signal to noise ratio improvement relative to that in $-100\,^{\circ}$C with respect
to CCD cooling temperature for a 300 second observation of a faint source in $1 \,\mu m$
band. The vertical dashed line indicates the temperature with the maximum signal to noise
ratio improvement. The QE of the chip is assumed to be linear to the temperature. 
The dark level shown in Fig.~\ref{darktemp}, and the a sky background value shown in 
Table \ref{tbl4} was used in the calculation.
}
\label{sn}
\end{figure*}

Fig. \ref{sn} shows the improvement in the signal to noise ratio (S/N) for a sky-limited 
observation of a faint point source ($> 20$ AB mag) with long exposures ($\sim 300$ seconds). 
Here we assume an aperture magnitude of a point source with the seeing of FWHM = $1\arcsec$ and a circular 
aperture size of $1.7\arcsec$ diameter. The background level is assumed for the bright night 
(6.2 DN/sec/pixel for $Y$-band). Once the object flux is much less than the sky noise, the result
does not depend on the exact value of the object flux. The QE of the camera at $1 \,\mu m$
is set to increase linearly with respect to its temperature as follows. The equation is
derived by utilizing the data from the manufacturer about the increase in the relative
QE values as a function of temperature for a typical deep depletion device:

\begin{displaymath}
QE(T) / QE(-100) = 2.003 + 0.0099T
\end{displaymath}
where $QE(-100)$ denotes the QE at $-100\,^{\circ}$C which is measured to be 0.2 from the
camera manufacturer's brochure. 
The dark current level is assumed to be as shown in Fig. ~\ref{darktemp}, and sky count of 
6.8 DN/sec/pixel in $Y$-band, respectively.
 Observation efficiency improves by nearly 12 \% and reaches its peak at around $-67\,^{\circ}$C 
due to the increase in QE as a function of the temperature but degrades
rapidly due to larger dark current level at higher temperatures. Hence, we set the operating
temperature of the camera to $-70\,^{\circ}$C.

\subsection{Flat field\label{flatsection}}

 On each clear night, we obtained twilight flat images at the start and 
the end of the observation. The exposure times were longer than
five seconds to avoid any shutter pattern (See Section~\ref{shutter}).
These flat images are analyzed to understand how stable the flat fields are.

\begin{figure*}
\epsscale{1.7}
\plotone{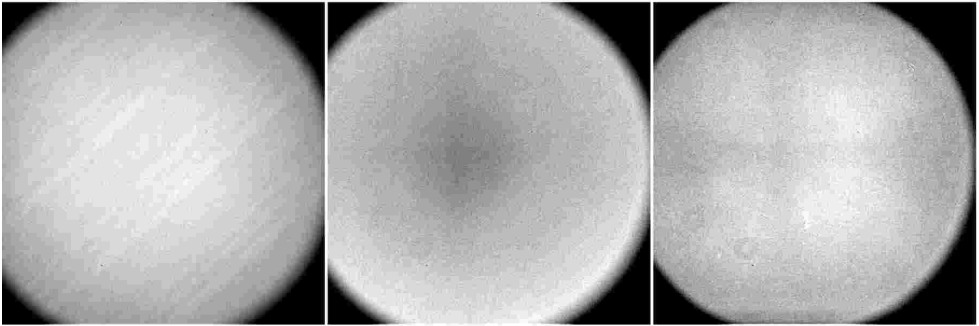}
\caption{Examples of flat images for $g$-band (right), $r$-band (middle), and $Y$-band (left).
All flat images show vignetting due to the focal reducer and the filters at all four corners.
For $g$ and $r$-bands, flat images show a inclined line patterns.
Flat images in other band look very similar to $Y$-band flat image.}
\label{exflat}
\end{figure*}

 Examples of flat images in $g$, $r$, and $Y$-bands are shown in Fig. \ref{exflat}.
  Vignetting can be seen in all these flat images at four corners of the chip,
 which is caused by a limited size of the focal reducer and the filters as described in
 Section \ref{freducer}. We note that the vignetting
 pattern changes from run to run, depending on how the focal reducer was aligned
 to the camera at the start of each run. Therefore, we recommend the users to
 use flat field images taken in the same run.
 Note that the $g$ and $r$-band flat images show diagonal line patterns across the whole
 area. And the $r$-band flat images show stronger tendency of limb brightening than in all the other
bands. These patterns in $g$ and $r$-band images may arise from the degraded AR 
coatings of the focal reducer lenses which are optimized for wavelength band of $0.8 - 1.1 \, \mu m$.
Flat images in all the other bands look similar to the $Y$-band flat image shown
 in the right panel of the Fig. \ref{exflat}, such that they show faint cross pattern across the
image. Although the origin of each pattern is not known, most of these patterns are
removed in the final flattened images.

\begin{figure*}
\epsscale{1.0}
\plotone{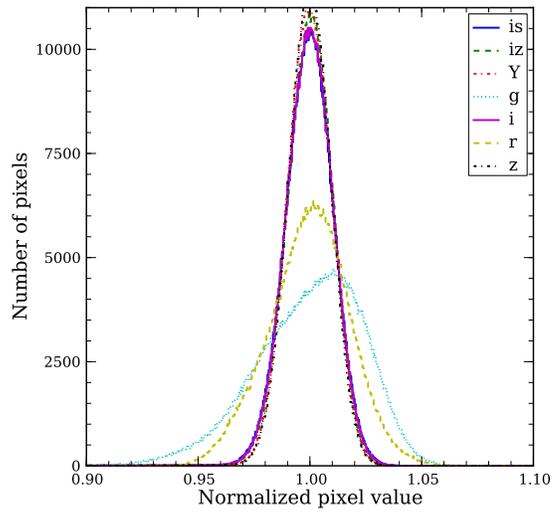}
\caption{Distributions of normalized pixel values in the flat images for each filter. Note
vignetted regions are not included in the distribution.}
\label{flatdist}
\end{figure*}

  Fig. \ref{flatdist} shows the distributions
 of normalized pixel values of flat field images inside a square box that excludes
 four vignetted regions. Ten twilight flat frames with pixel values higher than 10,000 ADU were used
 for the flat field construction in each band, and the expected level of sky noise is about $1 - 2 \,\%$.
  The figure shows that the pixel sensitivity variation is of order of a few percent for the filters
 other than $g$ and $r$, meaning that typical pixel-to-pixel sensitivity variation is very small.
  As for the $g$-band flat image, the pixel value distribution shows a peak higher
 than the normalized value due to the bright diagonal line patterns across its image. Similarly,
 the $r$-band flat shows a broader pixel value distribution since it shows a strong tendency
 of limb brightening than other band flats.

\begin{figure*}
\epsscale{1.6}
\plotone{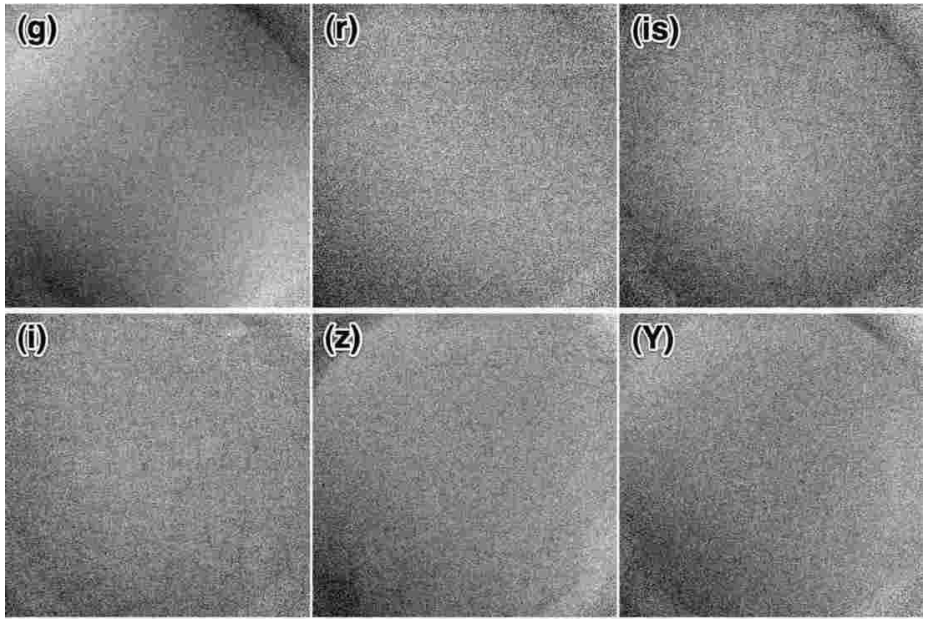}
\caption{Images of twilight flats taken on 2011 Aug. 28 divided by the ones taken on 2011 Aug. 30.
In all the images vignetting pattern is seen clearly at image corners. Gradient from lower
left to upper right direction is seen in this particular example sets but patterns vary on
different nights. See Section \ref{flatsection} for detailed explanation.}
\label{flatdiv}
\end{figure*}

  We compared the flat field images on four different nights during a single observing
 run to see how much the flat field patterns vary from night to night. Fig.~\ref{flatdiv} shows examples of
 a flat image from one night divided by another flat image taken on a different night.
 If there were no variation among flats, the divided images would be
 very flat. However, it can be seen that
 vignetting pattern is clearly visible at the edge of the image, and a gradient is seen
 across the image, of the peak to peak variation is as large as $\sim 1.5\,\%$ in some cases.
  Similar patterns are seen among the evening flat images divided by the morning flats
 taken on the same nights on some nights. 
 The variation of patterns is caused by the very small alignment change of
 focal reducer and the position change of the filters.  Observers have to pay special
 attention in setting up the instrument properly at the beginning of each run/night.
 Based on these
 results, observers are recommended to use the twilight flat images taken at the same night.

\begin{figure*}
\epsscale{1.7}
\plotone{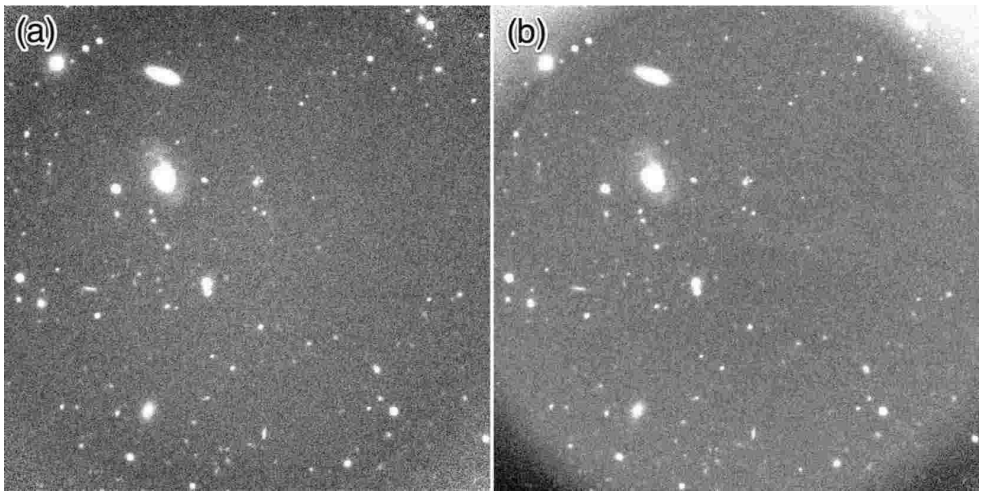}
\caption{$i$-band image (1800 seconds, total integration) flattened (a) with twilight flat, and  (b)
with dome flat. All the images including the flats were taken at the same night. See
Section \ref{flatsection} for the detailed explanation on the difference of dome flat and sky flat.
}
\label{skydome}
\end{figure*}

\begin{figure*}
\epsscale{1.0}
\plotone{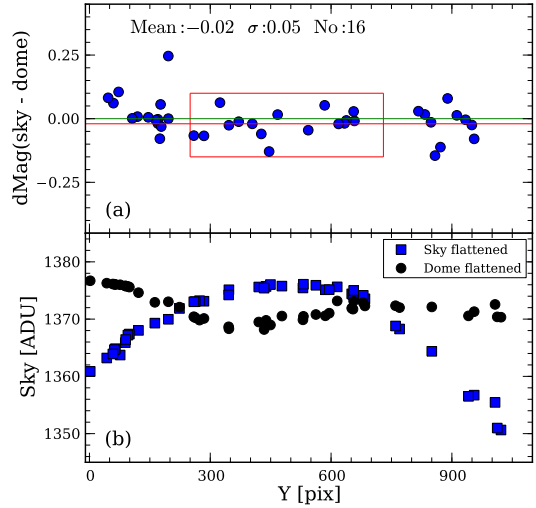}
\caption{(a) Magnitude difference between the photometry of the samples which
S/N is greater than 10 on both images shown in Fig.~\ref{skydome}, along the
Y axis of the image chip. The red box indicates the
samples used in derivation of mean value written in the upper part of the panel. The
samples near the both ends of Y axis are not used since all of them are located
in vignetted region. 
(b) Sky background around the sources detected in the central columns ($\mathrm{X}=400-600$),
along the Y axis of the image.}
\label{skydomephot}
\end{figure*}

Fig.~\ref{skydome} shows the stacked images of six 300 second $i$-band exposures
each flattened with a twilight flat image and a dome flat image taken on the same night.
The twilight flat images were obtained with 5 second exposure, and the dome flat images
with 5.5 seconds, to avoid the shutter pattern while securing enough high signals.
Due to the curvature of the dome, the illumination of the dome flat light 
turns out to be slightly brighter in the lower part than the other half. 
We ran SExtractor \citep{ber96} on two images to see the difference in the photometry,
which is shown in panel (a) of Fig.~\ref{skydomephot}. We did not find any
significant difference between two results for stars which S/N is greater than
10. No change in magnitude difference is seen along the vertical axis, either. There
are large scatters in upper and lower end of the chip (Y $\le 250$ or Y $\ge 750$),
but all of them are found to be in the vignetted region. The difference
in sky background between two images turned out to get larger in the outskirts
than in the central region, which was found to be $0.7 - 1.5 \,\%$ at the edge.
Since edge region is not recommended for photometry due to vignetting, the sky
difference would not be an important issue in practice.

\subsection{Gain and readout noise}

 The system electronics gain and its readout noise can be measured from its
image. From a pair of flat images and bias images, both quantities can be
measured with the following formulae \citep{bir06}:

\begin{displaymath}
\mathrm{gain} = {(\bar{F_{1}} + \bar{F_{2}}) - (\bar{B_{1}} + \bar{B_{2}}) \over
        {( \sigma^{2}_{F_{1} - F_{2}} ) - (\sigma^{2}_{B_{1} - B_{2}}}) }
\end{displaymath}

\begin{displaymath}
\mathrm{Readout\,\, noise} = {1 \over \sqrt{2}} \cdot \mathrm{gain} \mathit \cdot (\sigma_{B_{1} - B_{2}})
\end{displaymath}

where $\bar{F_{1}}$ and $\bar{F_{2}}$ are mean pixel values of the two flat images,
and $\bar{B_{1}}$, $\bar{B_{2}}$ the mean pixel values for two bias images.
$\sigma^{2}_{F_{1} - F_{2}}$ and $\sigma^{2}_{B_{1} - B_{2}}$ denote the variance
of the pixel values of the difference images, $F_{1} - F_{2}$ and $B_{1} - B_{2}$,
respectively.

Using a pair of $r$-band flats and bias images obtained during the 2010 August run,
we calculated the readout noise and the gain value of the camera.
 The gain and the readout noise are $1.0 \pm 0.1$ electrons/ADU, and 
$8.2 \pm 0.1$ electrons/pixel, respectively.

 We also checked whether the readout noise depends on the chip temperature, by
examining the standard deviation of mean pixel values of bias images taken under
four different temperatures in 2010 August run. The readout noise was measured
to be 8.3 electrons/pixel for images taken at $-40\,^{\circ}$C, and changed
to 8.2 as chip temperatures changed to $-50\,^{\circ}$C. But virtually no changes
were found for temperatures lower than that. Therefore we concluded that readout
noise is not sensitive to the chip temperature while bias level is.

Table \ref{tbl3} compares the measured values to those listed in the test report
by the camera manufacturer. The values are similar to the ones quoted
in the test report. 

\subsection{Shutter pattern\label{shutter}}

 The mechanical shutter in front of our science camera takes a finite time to open
and close when taking an image. This causes an imprint of the shutter pattern
 on the image when the exposure time is short.

\begin{figure*}
\epsscale{1.5}
\plotone{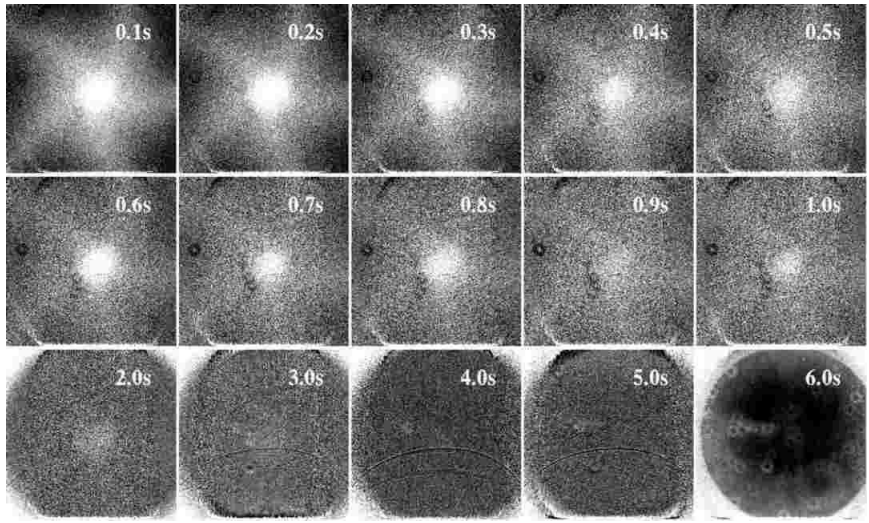}
\caption{Dome flat images of various exposures showing the shutter patterns. Shutter
pattern survives up to 3.0 second image.}
\label{shutterpattern}
\end{figure*}

 To examine how the shutter pattern affects an exposure as a function of
exposure time,
 we took a series of dome flats with different exposure times
 from 0.1 second to six seconds, which is shown in Fig. \ref{shutterpattern}.
 The five winged shutter pattern is clearly seen in the 0.1 second exposure and
it disappears as the exposures get longer.
 The shutter pattern remains in the exposures up to four second long, and it
does not appear in the longer exposure. The comparison of three second long exposure
and eight second long exposure reveals that the central region of the three second image is
about 0.4 \% brighter than the edge due to the remaining shutter pattern.
 From this result, it is recommended
to take flat images or science images with exposures longer than four seconds
if a precise photometry better than 0.4 \% is required. Or additional shutter
pattern images should be taken as a part of calibration observation.

\subsection{Linearity of the detector}

 As in other CCD imagers, our science camera shows non-linear behavior for the illumination near
its full-well capacity. To find out the linearity range of CQUEAN detector, we
obtained the images of the constant dome flat light at different exposure times.
 The experiment was repeated several times to check the stability
 of the non-linearity behavior.

\begin{figure*}
\epsscale{1.0}
\plotone{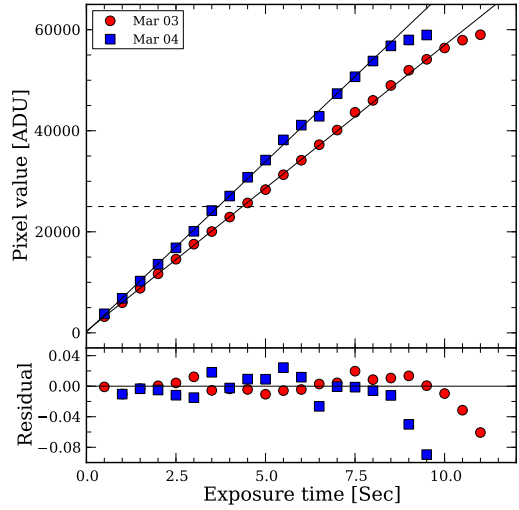}
\caption{Upper panel shows the CCD readout values against the exposure times of
constant dome flat light.
The dashed horizontal line indicates the upper limit of the readout values used to
fit the linear relations shown as solid lines in the figure. The lower panel
shows the residuals between the measured value and fitted line with respect to
measured value, against the exposure times.}
\label{linearity}
\end{figure*}

The upper panel of Fig. \ref{linearity} shows the readout values against different exposure times
from the experiments carried out over two nights in March 2011.  Since the dome flat
light brightnesses were different from each other on two occasions, the resultant
linearity curves show different slopes. However, they show similar patterns.
In both experiments, CCD readout values are more or less linear up to $\sim 50000$ ADUs.
The linearity curves show non-linear behavior beyond this range. 
The pixel values show a little variable behavior on each occasion. 

\subsection{Example of a reduced CQUEAN image and the fringe pattern}

 The data reduction of CQUEAN can follow the standard procedures of
bias and dark subtraction, followed by flat-fielding. For flat-fields,
we recommend twilight flats, although dome flats can be used if the
required photometric accuracy is not better than 0.02 mag.
 Also, one should be cautious about the temporal variation in the
 bias level. We recommend the users to take bias images just before
 and after the observation of each target and use these biases during
 the data reduction.

\begin{figure*}
\epsscale{1.6}
\plotone{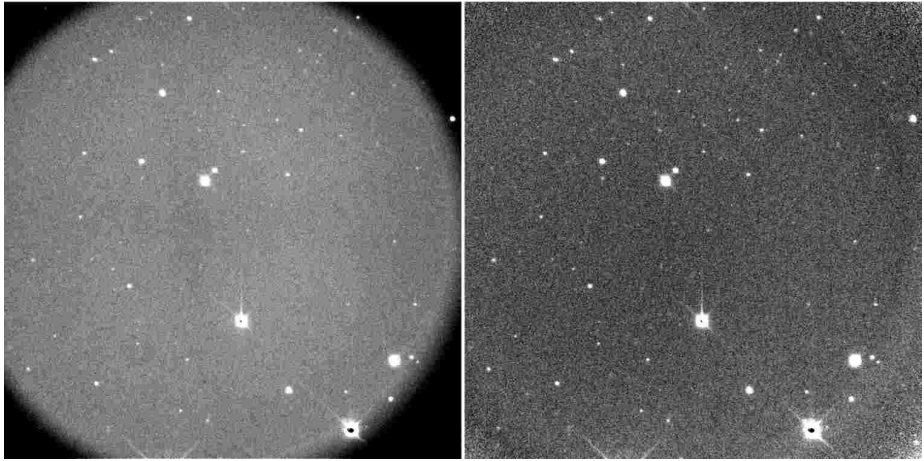}
\caption{Examples of a $z$-band raw CQUEAN image (left) and a reduced one (right). Note that
fringe pattern is not seen on both images. Although most of the system footprints
are removed in the reduced image, hint of vignetting is still seen at its corners.}
\label{ex_processing}
\end{figure*}

Fig. \ref{ex_processing} shows an example of a raw and a reduced CQUEAN images in $z$-band.
Fringe pattern is caused by the interferences of the light reflected on the
surface and inner silicon layer of the CCD chip, and is usually very severe
in the image of red optical bands taken with backside-illuminated, thinned
CCD \citep[e.g., see][]{im10a,jeo10}. However, no fringe patterns were noticed
in stacked CQUEAN images so far up to total integration time of two hours.
  We note that the standard reduction procedures removes most of the artifacts
 in the dark and flat-field, although the correction for the vignetted regions
 is not good.

\section{ON-SKY PERFORMANCE}

\subsection{Optical performance \label{section_psf}}

\begin{figure*}
\epsscale{2.0}
\plotone{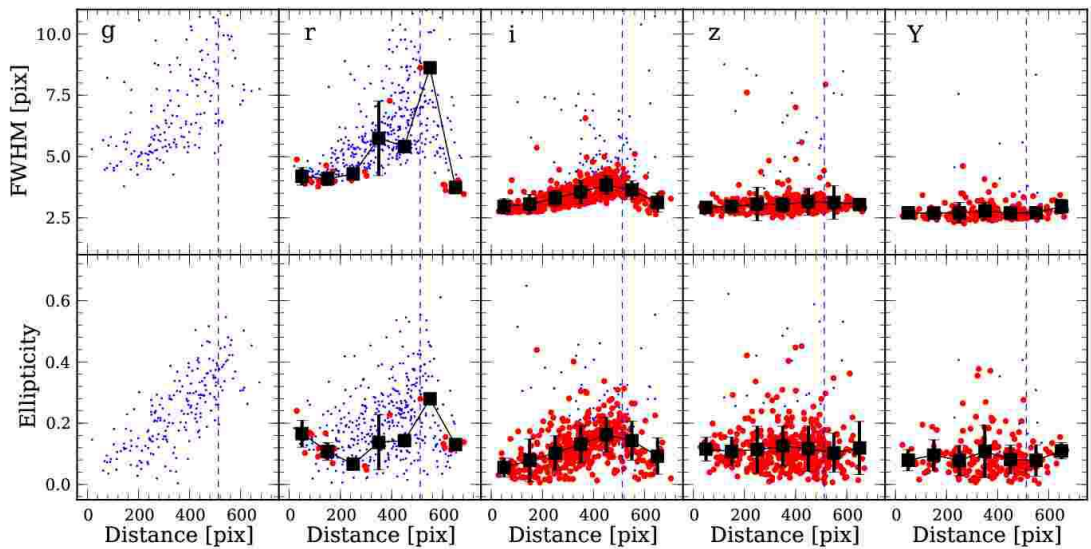}
\caption{FWHM (upper panels) and ellipticity (lower panels) of detected
sources in NGC 6633 images taken with CQUEAN. Vertical dashed line in
each panel indicates the range where vignetting is not present. The blue dots represent
all sources detected in the images, and red filled circles denote
the ones classified as point sources by SExtractor. Black squares represent
the mean values for the sources at every 100 pixel interval from the
CCD array center. 1-$\sigma$ ranges are indicated with vertical error bars.}
\label{fwhm}
\end{figure*}

 To estimate the optical performance of CQUEAN,
we examined the stellar image profiles in five band images. 
For the test, we obtained the images of an open cluster, NGC 6644, during the 2010
August run. SExtractor was run on the raw images to obtain
the image profile parameters such as FWHM, and ellipticity. Sources with
stellarity $\ge 0.95$ were regarded as point sources and their FWHM and
ellipticity were examined against their distances from the image center. Fig. \ref{fwhm}
displays the results for all sources including point sources.
 The optical performance of the focal reducer is optimized
for the wavelength band between $0.8 - 1.1 \, \mu m$,
 therefore the sources on the $g$ and $r$-band images have
 much larger FWHMs than on the other band images and show coma aberration: their
 FWHMs get larger and their image profiles
become more elongated as they are further from the image center. The coma is corrected 
for longer wavelength bands and it cannot be seen for the $z$ and $Y$-band images.
 This result indicates that while most region of the chip can be
used for photometry of faint sources for wavelength bands longer than $i$-band,
one should be cautious about using the outer part of the $g$ or $r$-band images
(radius $ \ge 200$ pixels).
 On short exposure frames under excellent seeing, we recorded $0.6\arcsec$ FWHM for
 point sources, demonstrating the good optical performance of the focal reducer.

\subsection{CQUEAN sensitivity}

We estimated the sensitivity of the CQUEAN science camera, from the data of
BD+47 0478 \citep{fuk96}, one of fundamental standard stars for SDSS, obtained
during the 2010 August run. The sensitivity of the system depends on the throughput
of the optical components and the QE of the detector. 
Table \ref{tbl4} lists the limiting magnitude during the bright time (full moon
illumination) and the dark time, for one hour total integration with 5-$\sigma$
error.

 The listed sky count rate is helpful when one wants to observe their targets
 in the sky-limited regime. Since the dark current is only 0.3 ADU/sec/pixel,
 the dominant source of the noise 
 is either the sky background noise or the readout noise in case of faint targets.
Considering the readout noise of 8.2 ADU and sky brightness in Table \ref{tbl4},
observations with exposure time longer than $5-15$ seconds are limited by the
sky background noise.

\begin{figure*}
\epsscale{1.0}
\plotone{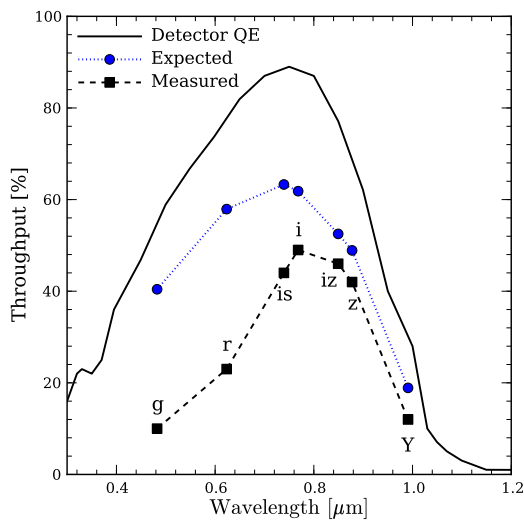}
\caption{Measured overall throughput of CQUEAN system. The solid line represents
the QE of the science camera chip alone, and the dotted line with circular points
the expected throughput from the combination of those for the chip QE, camera window, 
filters and two mirrors in the telescope. The dashed line with square points
represents the measured overall throughput.
}
\label{measured_thruput}
\end{figure*}

The expected overall throughput of CQUEAN can be calculated with data listed in
Table~\ref{tbl5}, and compared with actual measurement values in Fig.~\ref{measured_thruput}.
%
The expected value takes into account of the 
reflectivity of the two telescope mirrors, transmittance of the four lenses
in focal reducer and the camera window, and the QE of the CCD chip. 
Note that we could not include the wavelength dependent throughputs of the AR 
coating on the surfaces of the focal reducer lenses which had not been provided
by the manufacturer.
The actual overall throughput turned out to be $80-85$ \% of what is expected 
in $i$-band to $z$-band. The throughput is especially lower at shorter
wavelength bands: 25 \% at $g$, and 40 \% at $r$-band. 
The reduced performance in these two bands is caused by the
AR coating of focal reducer lenses optimized for longer
wavelength bands.

\subsection{Cosmic ray hits}
Deep depletion CCD is known to be more susceptible to cosmic ray hits than the back illuminated
thinned CCDs. Therefore we tried to estimate how much the cosmic ray hits can affect
our science mission with CQUEAN.
%
To get a general idea of how severe the cosmic ray hits might be on a single image, 
we identified cosmic ray hits with \verb|L.A.Cosmic| \citep{van01}, a cosmic ray
removal program for a single frame image, on 73 frames of 300 second exposure taken
with $g, z$, or $Y$-filter
during a CQUEAN run. We found that similar number of pixels are hit by the cosmic
rays among all filter images: $83.1 \pm 37.1$ for $g$, $118.6 \pm 42.5$ for $z$, and $110.6 \pm 40.9$
pixels for $Y$-band images. These values correspond to about 0.01 \% of total area. 
The affected pixels are easily removed during the stacking of multiple frames or with
other cosmic ray removal programs. Therefore we concluded that cosmic ray hits
do not affect our scientific mission of CQUEAN.

\section{SCIENTIFIC CAPABILITY OF CQUEAN}

  Since 2010 August, CQUEAN has been used to obtain photometric data for many
 scientific applications. 
Scientific targets include  quasar candidates at $\mathrm{z}=5.0 - 5.7$,
 transients such as Gamma Ray Bursts (GRB), supernovae, and variable young stellar objects,
 and host galaxies of bright quasars.
  Here, we summarize 
some of these observations briefly, in order to demonstrate the CQUEAN scientific capabilities.

\subsection{Stephan's Quintet}
\begin{figure*}
\epsscale{1.2}
\plotone{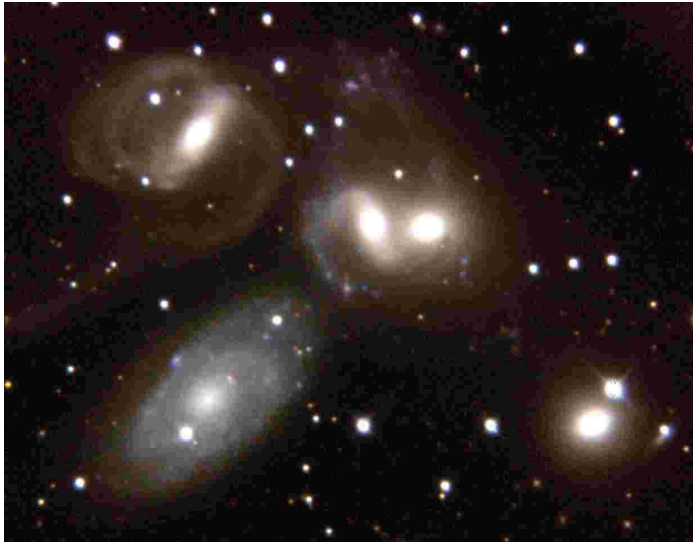}
\caption{Composite color image of Arp 319 made from $r$, $i$, and $z$-band images
taken with CQUEAN. Some rows in upper part of image were trimmed in the final product.
A faint arc line remains in the lower part of the image after the reduction, which is
due to inappropriate vertical shift speed setting of the camera.}
\label{arp319}
\end{figure*}

 Multi-band images of the Stephan's Quintet was obtained as a part of
 the science verification in our first run. Fig.~\ref{arp319} shows $r-i-z$
 color composite image of Stephan's Quintet. The image shows the expected
 optical performance of CQUEAN.

\subsection{Transient observation}

\begin{figure*}
\epsscale{1.0}
\plotone{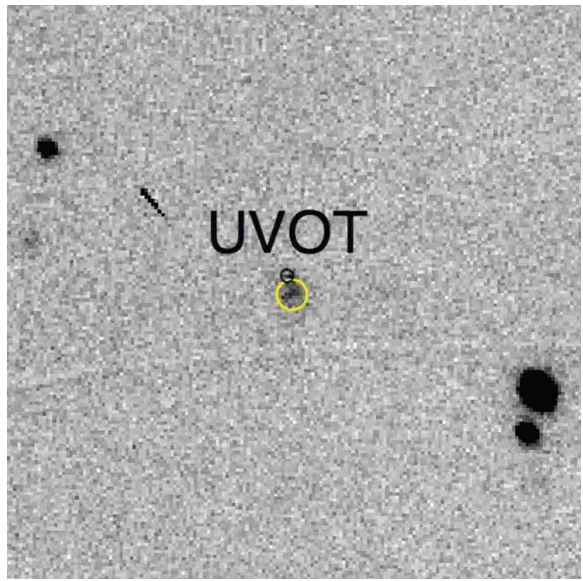}
\caption{Portion of $i$-band image of GRB 100816A and its host galaxy taken by CQUEAN.
Black circle shows the position of GRB 100816A afterglow detected by UVOT onboard Swift, with
which radius denotes its uncertainty. The extended emission (marked with yellow circle)
just below the UVOT position comes from the host galaxy of GRB 100816A.}
\label{grb100816a}
\end{figure*}

\begin{figure*}
\epsscale{2.0}
\plotone{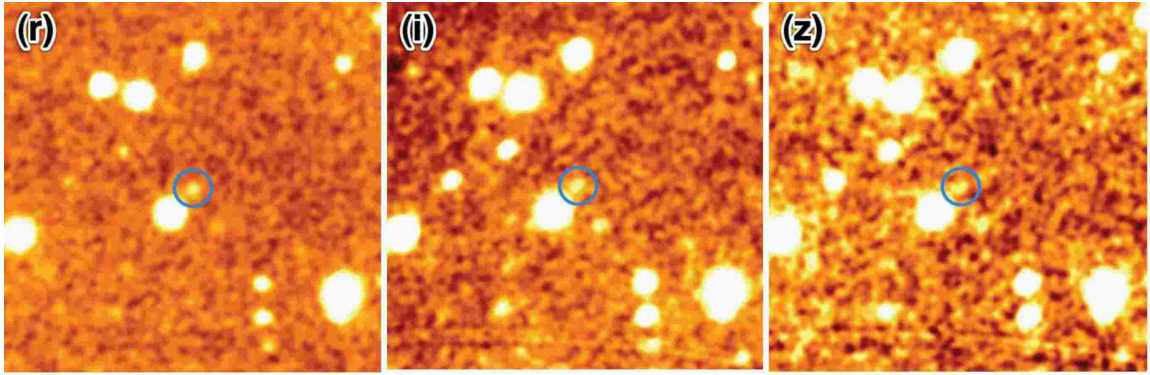}
\caption{Images of GRB 101225A taken by CQUEAN with $r$ (left), $i$ (middle), and $z$ (right)
filters. The field of view of each image is about $1.1\arcmin \times 1.1 \arcmin$, and the
target is located within the blue ring in the center. All images were smoothed to enhance
the contrast.}
\label{grb101225a}
\end{figure*}

 CQUEAN has been used to observe afterglows of GRBs \citep{im11, im10b, im10c, jeo11, par10}
  as a part of our GRB afterglow observation
 project \citep[e.g.,][]{lee10}. Most of the GRBs were
 observed with $r$, $i$, and $z$-band, and in some cases with $Y$-band.
 Upon the observation, preliminary photometry was
 reported to the GCN circular, and for some targets, the final photometric results
 were contributed to collaborative GRB studies. 
 We are also carrying out a long-term
 monitoring of Swift J1644+57 \citep{bur11} in multi-wavelength bands with many instruments
 including CQUEAN \citep{im11}, to understand the
 evolution of stellar remnants near a supermassive blackhole.
  Fig.~\ref{grb100816a} shows an $i$-band image of GRB 100816A taken about eight hours after its
 detection by Swift \citep{geh04}. The total integration time of the image is 2400 seconds.
 From this image, we were able to separate the GRB afterglow and
 its host galaxy which is about $i=20.58$ AB magnitude and located only
 $1.3\arcsec$ from the afterglow \citep{im10b}.
 Another image of a GRB afterglow is shown in Fig.~\ref{grb101225a},
 where we present $r$, $i$, and $z$-band images of GRB 101225A \citep{par10}
The CQUEAN data were used to characterize
 the early evolution of the optical afterglow which turn out to be dominated by
 a black-body radiation \citep{tho11}.

\subsection{High redshift quasar candidates}

\begin{figure*}
\epsscale{1.2}
\plotone{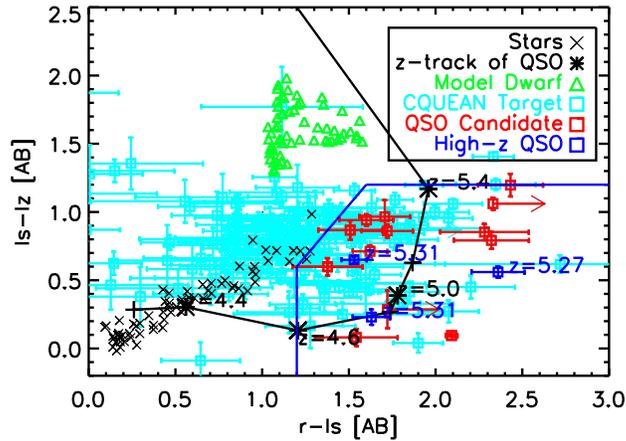}
\caption{Color-color diagram of the various targets. $is$ and $iz$ magnitudes
were obtained with CQUEAN, while $r$ magnitude was obtained from SDSS catalog.
The black line shows the track of quasars as a function of their redshifts.
The lower right region designated with blue lines indicates the color selection criteria for initial quasar
candidate samples. Note that previously known quasars, shown as blue open squares, all fall on
the selection region. The final quasar candidates, shown with red open squares,
were selected 
from this color criteria in combination with MIR colors.}
\label{qsoccd}
\end{figure*}

 We have been performing a survey of quasars at $5.0 < \mathrm{z} < 5.7$ using
 two custom designed filters.
 About 1000 targets were
 observed with the custom $is$ and $iz$-bands, among which about 20 were
 identified as promising candidates
 based on their colors from CQUEAN and mid infrared (MIR) colors of Wide-field Infrared Survey Explorer
 (WISE) \citep{wri10} data.
 Fig.~\ref{qsoccd} shows the locations
 of those candidates on the color-color diagram along with other targets
 observed.

\section{SUMMARY}

We developed CQUEAN, an optical CCD camera system which has an enhanced sensitivity
at $\sim 1 \,\mu m$ compared to the conventional CCD camera.
CQUEAN consists of a science camera, seven filters, a focal reducer
to enlarge its field of view, and a guide camera system.
CQUEAN also includes its in-house GUI software for controlling the instrument.
It is attached at the Cassegrain focus of the 2.1 m telescope at McDonald Observatory, USA.

The analysis of the CQUEAN data reveals that the camera is stable
and it meets the requirements set by the scientific purpose. 
With the focal reducer of our own design, the camera
produces images of the seeing as good as $0.6\arcsec$ without any fringe pattern in longer
wavelength bands, although coma aberration is present for $g$ and $r$-band
images as expected. We estimated the sensitivity of the camera to be
$z = 23.4$ AB mag. at one hour integration.
In addition, the characteristics of bias, dark, and flat images of the science
camera are examined, and the observation and data reduction strategy is discussed
to obtain better results under these system characteristics.
We also find that CQUEAN can obtain images of exposures up to 1200 second with its
auto-guider system.

  With the enhanced sensitivity at NIR together with a field of view of
 $4.8\arcmin \times 4.8 \arcmin$, CQUEAN has been serving
 as a useful instrument for the observation of red, astronomical objects such as
 high redshift quasar candidates, afterglows of GRBs, and supernovae.

\acknowledgments
This work is supported by the Creative Research Initiatives program of the Korea Science and Engineering Foundation (KOSEF), grant No. 2009-0063616, funded by the Korea government (MEST).
 This paper made use of the data obtained with a telescope
at the McDonald Observatory, TX, USA. We thank the staffs of McDonald Observatory, including David Doss and John Kuehne, for their assistance of CQUEAN observations, and P. Samuel Odoms for allowing us to use \textit{agdr}, the auto-guiding software. Authors also thank observers who helped obtain the data discussed in the paper.

\begin{deluxetable}{ll}
\tablecolumns{2}
\tablewidth{0pc}
\tablecaption{Specification of Andor iKon-M M934 BR-DD camera\tablenotemark{\dagger}}
\tablehead{}
\startdata
Type of CCD chip             & A deep-depletion chip made by E2V \\
Pixel number                 & $1024 \times 1024$ \\
Pixel size                   & $13\,\mu m \times 13 \,\mu m$ \\
Image area                   & $ 13.3 \,mm \times 13.3 \,mm $ \\
Pixel scale\tablenotemark{a}  & $ 0.281\arcsec$/pixel \\
Field of view\tablenotemark{a} & $4.8 \arcmin \times 4.8 \arcmin$\\
Active area pixel well depth & 100,000 e- (typical) \\
Output saturation            & 200,000 e- (typical) \\
Pixel readout rate           & 2.5, 1.0\tablenotemark{b}, 0.05 MHz \\
Array readout time           & $\sim0.5, 1, 20$ seconds \\ 
Pixel readout noise (typical)&  9.0, 7.0, 2.8 electrons \\
Digitization                 & 16 bit at all readout speeds \\
Cooling                      & Down to $-80\,^{\circ}$C with air-cooling \\
                             & Down to$-100\,^{\circ}$C with optional water-cooling\\
\enddata
\tablenotetext{\dagger}{Based on the brochure}
\tablenotetext{a}{With focal reducer}
\tablenotetext{b}{Current readout speed for CQUEAN}
\label{ccdspec}
\end{deluxetable}

\begin{deluxetable}{cccrrr}
\tablecolumns{6}
\tablewidth{0pc}
\tablecaption{Descriptions of the focal reducer.\label{tbl2}}
\tablehead{
\colhead{Surface\tablenotemark{a}} & \colhead{Description} & \colhead{Medium} & 
\colhead{Radius of} & \colhead{Thickness\tablenotemark{b}} & 
\colhead{Diameter\tablenotemark{b}} \\
\colhead{} & \colhead{} & \colhead{} & \colhead{curvature\tablenotemark{b}} & \colhead{} & \colhead{}
}
\startdata
1  &  Lens 1  &  N-BAF10\tablenotemark{c}  &   91.188  &  14.00  &  66.00 \\
2  &  Lens 2  &  N-SF66\tablenotemark{c}   & -118.153  &   4.00  &  66.00 \\
3  &  \nodata &  \nodata                   &  118.153  &   8.20  &  63.00 \\
4  &  Lens 3  &  SF57\tablenotemark{c}     &   84.927  &  11.00  &  66.00 \\
5  &  \nodata &  \nodata                   & Infinity  &  15.80  &  66.00 \\
6  &  Lens 4  &  N-FK5\tablenotemark{c}    &   43.774  &   7.00  &  50.00 \\
7  &  \nodata &  \nodata                   &   31.344  &  43.03  &  42.00 \\
8  &  Filter  &  Silica                    & Infinity  &   5.00  &  28.00 \\
9  &  \nodata &  \nodata                   & Infinity  &  23.58  &  28.00 \\
10  &  CCD    &  \nodata                   & Infinity  & \nodata & \nodata  \\
\enddata
\tablenotetext{a}{See Fig.~\ref{lens} for the surface numbers.}
\tablenotetext{b}{All values are in unit of millimeter.}
\tablenotetext{c}{Optical glasses manufactured by Schott.}
\end{deluxetable}

\begin{deluxetable}{lcc}
\tablecolumns{3}
\tablewidth{0pc}
\tablecaption{Characteristics of the CCD in the science camera\label{tbl3}}
\tablehead{
\colhead{} &
\colhead{Measured} &
\colhead{Reported\tablenotemark{\dagger}}
}
\startdata
Gain (e-/ADU)  & $1.0\pm0.1$  &  1.2 \\
Readout noise (e-/pixel) & $8.2\pm0.1$  &  7.0 \\
Dark level (e-/sec/pixel) & $0.23\pm 0.01$   & 0.1\\
Saturation level (e-/pixel) & $\sim57000$  & 75957 \\
Base level (ADU) & $4425\sim4487$ & 4632 \\
\enddata
\tablenotetext{\dagger}{Tested and reported by the camera manufacturer. The camera
 was cooled down to $-80\,^{\circ}$C for the measurement.}
\end{deluxetable}

\begin{deluxetable}{cccccc}
\tablecolumns{6}
\tablewidth{0pc}
\tablecaption{CQUEAN Point-source sensitivity\tablenotemark{a}\label{tbl4}}
\tablehead{
\colhead{Condition} &
\colhead{Filter} &
\colhead{$\lambda_{eff}$} &
\colhead{Sky count} &
\colhead{Sky brightness} &
\colhead{Magnitude limit\tablenotemark{b}} \\
\colhead{} &
\colhead{} &
\colhead{($\mu m$)} &
\colhead{(DN/sec/pixel)} &
\colhead{(mag/arcsec$^2$)} &
\colhead{(AB mag)} 
}
\startdata
       &$r$   & 0.623 &  1.2 & 21.3 & 24.1 \\
       &$i$   & 0.768 &  4.6 & 20.6 & 24.3 \\
Dark   &$is$  & 0.739 &  5.0 & 20.3 & 24.0 \\
time   &$iz$  & 0.849 &  8.0 & 19.8 & 23.8 \\
       &$z$   & 0.877 &  5.2 & 19.6 & 23.4 \\
       &$Y$   & 0.991 &  3.8 & 18.5 & 22.1 \\
\hline
       &$r$   & 0.623 &  4.1 & 19.8 & 23.4 \\
       &$i$   & 0.768 & 11.1 & 19.7 & 23.8 \\
Bright &$is$  & 0.739 &  8.4 & 19.7 & 23.7 \\
time   &$iz$  & 0.849 & 11.7 & 19.3 & 23.5 \\
       &$z$   & 0.877 &  8.4 & 19.1 & 23.1 \\
       &$Y$   & 0.991 & 6.2  & 17.9 & 21.8 \\
\enddata
\tablenotetext{a}{Based on the data obtained on 2010 August run}
\tablenotetext{b}{With total one hour integration and 5-$\sigma$ error}
\end{deluxetable}

\begin{deluxetable}{cccccc}
\tablecolumns{6}
\tablewidth{0pc}
\tablecaption{The throughput estimation for CQUEAN\label{tbl5}}
\tablehead{
\colhead{Filter} & \multicolumn{4}{c}{Expected} & \colhead{Measured} \\
\cline{2-5}\\
\colhead{} & \colhead{Telescope\tablenotemark{a}} & \colhead{Focal reducer\tablenotemark{b}} & 
  \colhead{Camera\tablenotemark{c}} & \colhead{Combined} & \colhead{} \\
}
\startdata
$g$   & 0.840 & 0.934 & 0.52 & 0.40 & 0.10 \\
$r$   & 0.820 & 0.984 & 0.72 & 0.58 & 0.23 \\
$is$  & 0.780 & 0.991 & 0.82 & 0.63 & 0.44 \\
$i$   & 0.762 & 0.991 & 0.82 & 0.62 & 0.49 \\
$iz$  & 0.783 & 0.991 & 0.72 & 0.53 & 0.46 \\
$z$   & 0.766 & 0.991 & 0.64 & 0.49 & 0.42 \\
$Y$   & 0.861 & 0.993 & 0.22 & 0.19 & 0.12 \\
\enddata
\tablenotetext{a}{Effective throughput of two mirrors, based on the reflectivity measured on 2011 May \citep{dos11}}
\tablenotetext{b}{Effective throughput of four lenses, based on the generic trasmittance 
  for the lens material used in the focal reducer. Note that the throughput of the AR coatings
  on the focal reducer lenses are not included.}
\tablenotetext{c}{Effective throughput of camera window and CCD QE, based on the generic data.}
\end{deluxetable}

\clearpage

\clearpage

\clearpage

\clearpage

\clearpage

\clearpage

\clearpage

\clearpage

\clearpage

\clearpage

\clearpage

\clearpage

\clearpage

\end{document}